\documentclass[conference]{IEEEtran}
\IEEEoverridecommandlockouts
\usepackage{cite}
\usepackage{amsmath,amssymb,amsfonts}
\usepackage{algorithmic}
\usepackage{graphicx}
\usepackage{textcomp}
\usepackage{xcolor}
\usepackage{float}
\usepackage{booktabs} 
\usepackage{balance}

\def\BibTeX{{\rm B\kern-.05em{\sc i\kern-.025em b}\kern-.08em
    T\kern-.1667em\lower.7ex\hbox{E}\kern-.125emX}}
\begin{document}

\title{\huge Enhancing Anomaly Resilience in Research Networks:\\ 
A Large-Scale Forecasting Benchmark for \\
Dynamic Security Baselining}

\author{\IEEEauthorblockN{Mohammad Arafath Uddin Shariff and Byrav Ramamurthy}
\IEEEauthorblockA{\textit{School of Computing} \\
\textit{University of Nebraska-Lincoln}\\
Lincoln, NE, USA \\
mshariff2@nebraska.edu, ramamurthy@unl.edu}
}

\IEEEoverridecommandlockouts
\IEEEpubid{\begin{minipage}{\textwidth}\ \\[12pt] \centering
  \copyright~2026 IEEE. Personal use of this material is permitted. Permission from IEEE must be obtained for all other uses, in any current or future media, including reprinting/republishing this material for advertising or promotional purposes, creating new collective works, for resale or redistribution to servers or lists, or reuse of any copyrighted component of this work in other works.
\end{minipage}}

\maketitle

\begin{abstract}
Research and Education Networks (RENs) serve as critical infrastructure for scientific discovery, yet they face a unique security paradox: their normal traffic patterns which are characterized by massive, bursty ``elephant flows" are statistically indistinguishable from volumetric attacks such as DDoS to conventional monitoring systems. This similarity leads to high false-positive rates in anomaly detection, blinding security operators to genuine threats. In this paper, we propose and evaluate a high-fidelity traffic forecasting framework designed to establish dynamic security baselines for RENs. Leveraging an exclusive 57-day Internet2 dataset spanning ten backbone routers (13.7 billion packets), we perform the first large-scale benchmark of anomaly-aware forecasting models in this domain. We systematically evaluate six model families, from SARIMA to state-of-the-art long-sequence architectures (TiDE, PatchTST), across 960 experimental configurations. Our results demonstrate that these advanced architectures, particularly TiDE, reduce baseline prediction error by 30-42\% compared to traditional methods ($p < 0.001$), significantly improving the distinction between legitimate scientific bursts and potential anomalies. Furthermore, we introduce a novel anomaly-integration strategy that improves model robustness by 3.3\% in the presence of noise. This work provides the first statistically validated framework for distinguishing scientific workflows from network attacks, enabling more autonomous and resilient network security operations.
\end{abstract}


\section{Introduction}

Research and Education Networks (RENs), such as Internet2 and ESnet, form the nervous system of modern science, facilitating petabyte-scale data transfers for high-energy physics, genomics, and climate research \cite{monga2018sdn}. As critical infrastructure, these networks are high-value targets for cyber threats, necessitating robust, AI-driven cybersecurity frameworks for their protection \cite{ahamed2025ai} However, securing RENs presents a unique challenge: the ``normal" behavior of a REN often mimics the signatures of volumetric attacks. A single scientific workflow can saturate a 100 Gbps link for hours, generating extreme burstiness and traffic spikes that traditional Intrusion Detection Systems (IDS) might flag as a Distributed Denial of Service (DDoS) attack \cite{dart2013science}.

This ambiguity creates a severe operational security risk. If security operators tune detection thresholds too loosely, they miss genuine attacks (false negatives). If they tune them too tightly, legitimate scientific transfers trigger alarms (false positives), leading to ``alert fatigue" and potentially the automated blocking of critical research data. To resolve this, REN security requires a dynamic baseline—a highly accurate, predictive model of what ``normal" scientific traffic looks like at any given moment, enabling the isolation of true anomalies (attacks, failures) from predicted elephant flows.

Despite this need, current security baselining techniques in RENs remain rudimentary. Three fundamental barriers have hindered progress: (1) Data Scarcity: Real-world REN datasets are rarely accessible for security research due to privacy concerns \cite{nottingham2020challenges}; (2) Lack of Benchmarking: There is no systematic comparison of modern deep learning architectures for creating robust security baselines in these environments; and 

\IEEEpubidadjcol
(3) Anomaly Blindness: Existing forecasting models often fail to account for the presence of anomalies in training data, leading to poisoned baselines that normalize attack patterns.

In this work, we address these challenges by developing a robust, anomaly-aware forecasting framework for dynamic security baselining. We leverage an exclusive 57-day dataset from the Internet2 backbone to benchmark six model families, ranging from classical statistical methods to modern Transformers. By integrating unsupervised anomaly detection (Isolation Forest, Local Outlier Factor) directly into the forecasting pipeline, we demonstrate how to build models that are resilient to noise and capable of predicting legitimate bursts with high fidelity.

Our contributions to network security and measurement are: (1) We analyze 13.7 billion packets from ten Internet2 routers, providing rare insight into the baseline traffic patterns of critical scientific infrastructure; (2) We conduct the first systematic evaluation of advanced long-sequence architectures (including MLP-based TiDE and Transformer-based PatchTST) against traditional baselines for REN traffic, demonstrating that Transformers offer superior fidelity for security baselining; (3) We introduce and evaluate four strategies for integrating anomaly detection into the forecasting loop, showing that masking anomalies during training significantly improves the robustness of RNN-based security models; (4) We analyze the trade-offs between baseline accuracy and computational cost, providing actionable recommendations for deploying real-time anomaly detection support systems.

The remainder of this paper is organized as follows. Section II reviews related work in network security and traffic analysis. Section III details our dataset and the anomaly-aware methodology. Section IV describes the experimental design. Section V presents the benchmarking results. Finally, Section VI discusses the security implications and concludes the paper.

\section{Related Work and Positioning}

\subsection{Traffic Characteristics and Security Risks in RENs}
Research and Education Networks (RENs) exhibit traffic patterns that are fundamentally distinct from commercial WANs, creating unique challenges for security monitoring. Rogers and Christensen first demonstrated that Internet2 traffic lacks the self-similar, human-driven patterns of commercial networks, instead following heavy-tailed distributions where under 1\% of flows account for over 80\% of volume \cite{rogers2001fluid}. This operational reality is shaped by architectural patterns like the Science DMZ, which prioritizes throughput by bypassing traditional firewalls \cite{dart2013science}.

From a security perspective, this creates a significant ``masking" effect. As noted by Kiran and Chhabra, REN workloads are non-stationary and unpredictable \cite{kiran2019understanding}. This non-stationarity makes it difficult for threshold-based Intrusion Detection Systems (IDS) to distinguish between a legitimate 100 Gbps scientific transfer and a volumetric DDoS attack. Barford et al. analyzed such anomalies using signal processing, highlighting the difficulty of separating ``flash crowd" events from malicious attacks \cite{barford2002signal}. Recent work by Giannakou et al. at NERSC further underscores the complexity of data movement patterns in modern HPC environments, necessitating more adaptive baselining techniques \cite{giannakou2024understanding}.

\subsection{Anomaly Detection Challenges in High-Speed Networks}
Anomaly detection in high-speed networks is a well-studied field, yet significant challenges persist. Chandola et al. provide a comprehensive survey of techniques \cite{chandola2009anomaly}, but applying them to network intrusion detection is fraught with difficulty due to the ``semantic gap" and high cost of false positives \cite{sommer2010outside}. Gates and Taylor challenged the traditional anomaly detection paradigm, noting that ``normal" traffic is rarely static \cite{gates2006challenging}, a view supported by Mahoney and Chan who proposed learning non-stationary models for detecting novel attacks \cite{mahoney2002learning}.

Unsupervised methods are often required due to the scarcity of labeled attack data. Eskin et al. proposed a geometric framework for unsupervised detection \cite{eskin2002geometric}, a concept we operationalize using Isolation Forests \cite{liu2008isolation} and Local Outlier Factor (LOF) \cite{breunig2000lof}. While Lakhina et al. demonstrated the utility of subspace methods (PCA) for diagnosing backbone anomalies \cite{lakhina2004diagnosing}, our work advances this by integrating these unsupervised detectors directly into the forecasting pipeline to create a robust, dynamic baseline.

\subsection{Forecasting Models for Security Baselining}
Forecasting models serve as the foundation for dynamic anomaly detection. Foundational work by Papagiannaki et al. explored long-term forecasting on earlier backbones \cite{papagiannaki2005long}, but these models predate current petabyte-scale flows. Recent studies using SARIMA reported high error rates during scientific-scale bursts \cite{mohammed2021predicting}, while Liu et al. highlighted the inadequacy of classical baselines for capacity planning \cite{liu2018comprehensive}.

The advent of deep learning has led to more advanced architectures. A prior study by Shariff et al. demonstrated hybrid GRU-LSTM effectiveness on two routers \cite{shariff2025traffic}. More recently, Graph Neural Networks (GNNs) like DCRNN have shown promise for spatio-temporal modeling \cite{andreoletti2019network}, and systems like APRIL have begun to use forecasts for intelligent load balancing \cite{8737537}. However, a critical gap remains: these advanced long-sequence architectures, including modern MLP-based encoders like TiDE \cite{das2023long} and Transformers like PatchTST \cite{nie2022time}, have never been systematically benchmarked for their ability to serve as security baselines on a large-scale REN backbone. Table \ref{tab:comparison} summarizes this gap.

\begin{table}[!t]
\caption{Positioning Against Prior REN Forecasting and Security Studies}
\centering
\resizebox{\columnwidth}{!}{%
\begin{tabular}{lrrcl}
\toprule
\textbf{Study} & \textbf{Routers} & \textbf{Duration} & \textbf{Models} & \textbf{Anomaly} \\
\midrule

Mohammed \textit{et al.} \cite{mohammed2021predicting} & 1 & 30 days & SARIMA & None \\
Liu \textit{et al.} \cite{liu2018comprehensive} & 1 & 365 days & Statistical & None \\
Andreoletti \textit{et al.} \cite{andreoletti2019network} & 23 & $\sim$30 days & DCRNN (GNN) & None \\
Shariff \textit{et al.} \cite{shariff2025traffic} & 2 & 30 days & GRU-LSTM & None \\
\midrule
\textbf{This Work} & \textbf{10} & \textbf{57 days} & \textbf{6 families} & \textbf{4 strategies} \\
\bottomrule
\end{tabular}
}
\label{tab:comparison}
\end{table}

\section{Methodology and Dataset Analysis}

We leverage an exclusive 57-day Internet2 dataset containing 314 million flows and 13.7 billion packets from ten geographically distributed backbone routers, providing the largest, most diverse REN corpus for security baselining to date. Table~\ref{tab:dataset} summarizes router-level scale and coverage.

\begin{figure}[h!]
\centering
\includegraphics[width=1\linewidth,trim={9.4cm 2.3cm 3.8cm 5cm},clip]{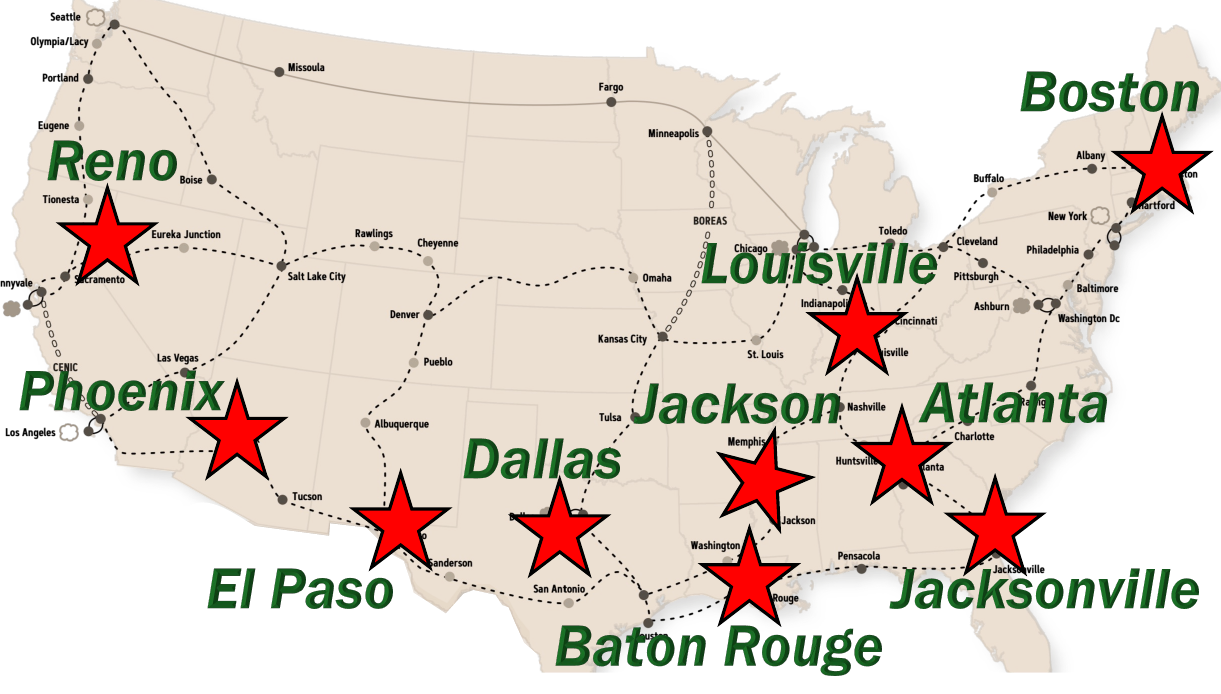}
\caption{Geographic distribution of the ten Internet2 routers. The diversity of this backbone dataset allows us to test security baseline generalizability across both high-volume hubs and smaller regional nodes.}
\label{fig:map}
\end{figure}

\begin{table}[!t]
\caption{Internet2 Traffic Corpus Summary by Router}
\centering
\begin{tabular}{lrrrr}
\toprule
\textbf{Router} & \textbf{Raw GB} & \textbf{Flows (M)} & \textbf{Packets (B)} & \textbf{Hours} \\
\midrule
Atlanta & 8.49 & 28.03 & 0.61 & 1369 \\
Dallas & 47.76 & 149.78 & 2.42 & 1322 \\
Baton Rouge & 28.43 & 80.06 & 1.29 & 1343 \\
Boston & 0.14 & 0.39 & 0.02 & 789 \\
El Paso & 3.33 & 9.36 & 0.32 & 1378 \\
Jackson & 4.79 & 13.40 & 0.23 & 1369 \\
Jacksonville & 6.95 & 19.43 & 0.70 & 1338 \\
Louisville & 1.12 & 3.15 & 0.28 & 1466 \\
Phoenix & 3.12 & 8.77 & 0.27 & 1369 \\
Reno & 12.66 & 35.50 & 0.47 & 861 \\
\midrule
\textbf{Total} & \textbf{147.7} & \textbf{314.0} & \textbf{13.7} & \textbf{12,604} \\
\bottomrule
\end{tabular}
\label{tab:dataset}
\end{table}

The dataset exhibits significant router heterogeneity: Dallas processes the highest volume (149.78M flows, 2.42B packets) while Boston handles the smallest load (0.39M flows, 20M packets). This 380$\times$ variation in flow volume and geographic distribution across major research institutions enables comprehensive model evaluation under diverse traffic conditions. Coverage ranges from 789 to 1,466 hours per router due to data availability windows, with all routers providing sufficient temporal scope for robust train/validation/test splits. 


While network-wide spatial dependencies exist, we focus on univariate, per-router modeling as it represents the standard deployment model for edge-based Intrusion Detection Systems (IDS). By establishing independent baselines for each core router, we ensure that security alerts are localized and actionable.

Our exploratory analysis reveals three key characteristics differentiating REN traffic from commercial networks: 1) Muted Diurnal Cycles, where traffic drops only $\sim$15\% overnight (Fig. \ref{fig:daily_curves}), reflecting machine-driven workflows; 2) Extreme Heavy-Tailed Distributions, with the top 1\% of flows accounting for $>80\%$ of volume; and 3) Striking Router Heterogeneity (Table \ref{tab:dataset}). This justifies our multi-site, per-router benchmarking approach.

\subsection{Scalable ETL Pipeline}

\begin{figure}[h!]
\centering
\includegraphics[width=1\linewidth, trim=9.5cm 0.25cm 9.5cm 0.1cm, clip]{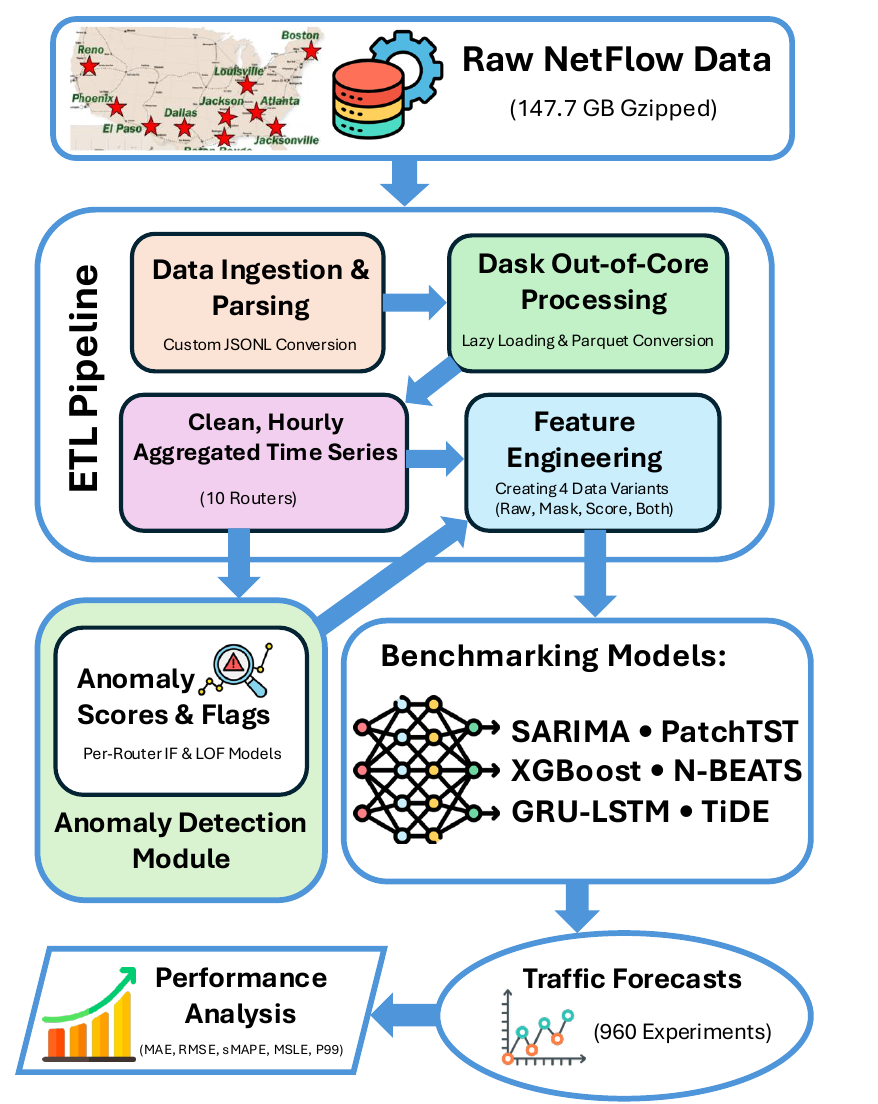}
\caption{The end-to-end architecture of our anomaly-aware forecasting framework. Raw Internet2 NetFlow data is processed by a multi-stage, Dask-based ETL pipeline to produce a clean, hourly aggregated time series. This clean data is then fed into an Anomaly Detection Module to generate scores and flags, which are combined in the Feature Engineering stage to create four distinct data variants. Finally, these variants are used in a comprehensive benchmark of six model families to generate 960 sets of experimental forecasts for performance analysis.}
\label{fig:etl_pipeline}
\end{figure}

Traditional in-memory data processing fails at this volume, a common challenge when extracting actionable insights from large-scale data systems \cite{begum2023comparative}. To overcome this, we engineered a distributed, Dask-based ETL pipeline, shown in Fig.~\ref{fig:etl_pipeline}. The pipeline first performs Data Ingestion \& Parsing to handle the raw, non-standard JSON fragments. The data then flows to the Dask Out-of-Core Processing stage, which uses lazy loading and Parquet conversion to manage the massive scale. This produces a Clean, Hourly Aggregated univariate Time Series by summing the \texttt{in\_packets} field for each of the 10 routers. This clean data serves two purposes: it is the input for the Anomaly Detection Module and is combined with the anomaly scores in the Feature Engineering stage to create the final four data variants used in our benchmark.

\begin{figure}[h!]
\centering
\includegraphics[width=1\linewidth]{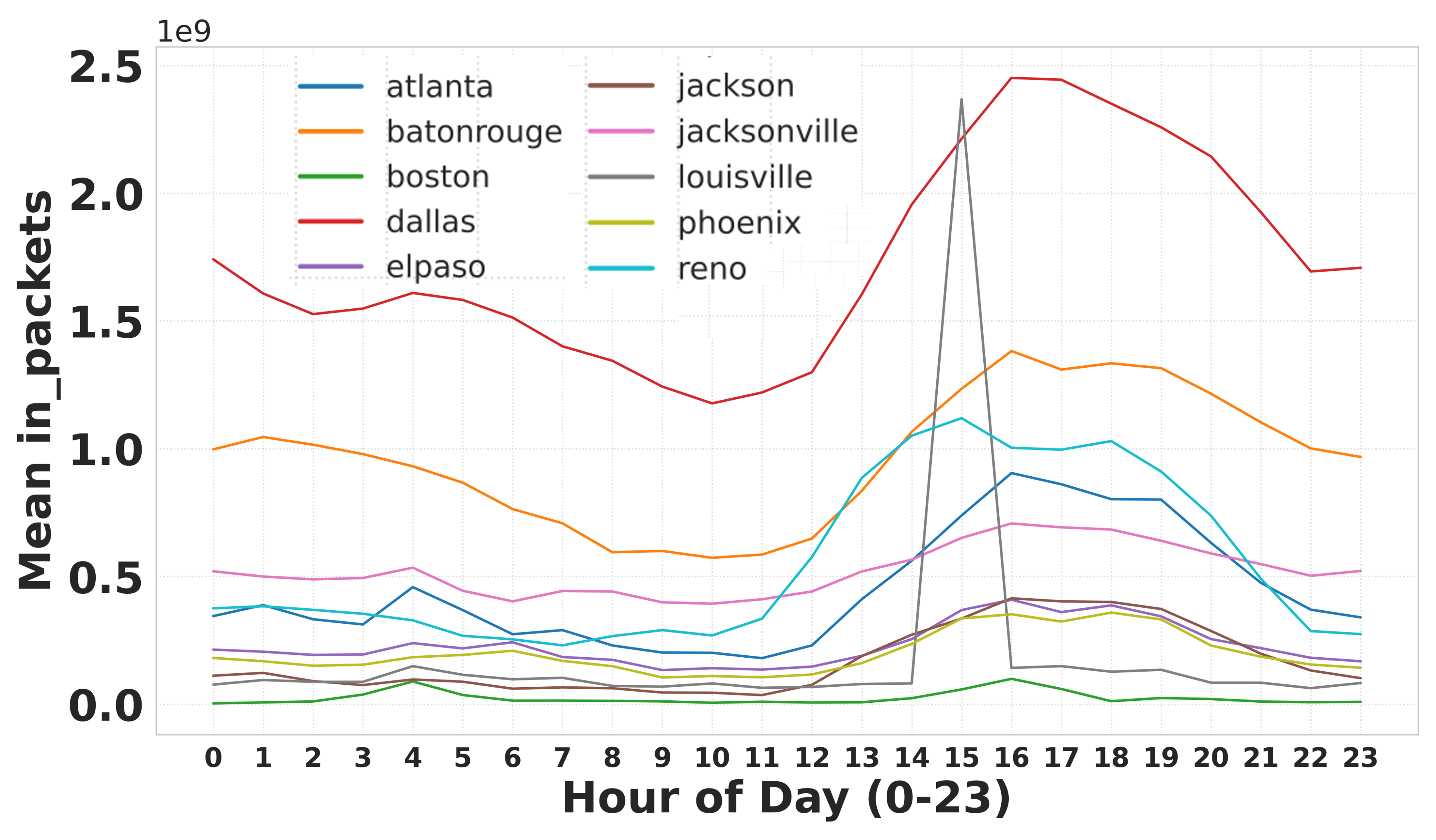}
\caption{Mean hourly traffic across ten routers. The muted diurnal cycles (only ~15\% variation) and extreme burstiness make REN traffic distinguishable from commercial traffic, complicating traditional DDoS detection thresholds.}
\label{fig:daily_curves}
\end{figure}

After parsing and hourly aggregation, we split the time series chronologically (70/15/15\%) for realistic forecasting. Figure~\ref{fig:daily_curves} illustrates muted diurnal patterns across all routers, with overnight traffic dropping only 15\% compared to peak hours, far less than the 50-70\% reductions typical in commercial WANs. This reflects the continuous, machine-driven nature of scientific workflows that operate independently of human activity cycles. 

To distinguish legitimate scientific bursts from operational anomalies and potential attacks, we deploy both Isolation Forest (IF) and Local Outlier Factor (LOF), trained per-router. In the context of REN security, we define an `anomaly' as any traffic pattern that significantly deviates from the learned baseline. Unlike commercial networks where anomalies are often malicious such as DDoS, REN anomalies are bimodal: they can be legitimate scientific `elephant flows' or actual security events. Our framework uses Isolation Forest not just to flag outliers, but to clean the training data, allowing the forecasting model to learn the true underlying baseline of legitimate traffic. This separation is critical for reducing false positives in downstream IDS.

Anomalous likelihood for IF is scored as:
\begin{equation}
s(x,n) = 2^{-\frac{E(h(x))}{c(n)}}
\label{eq:if_score}
\end{equation}
and for LOF as:
\begin{equation}
LOF_k(x) = \frac{\sum_{y \in N_k(x)} lrd_k(y)/lrd_k(x)}{|N_k(x)|}
\label{eq:lof_score}
\end{equation}
where $E(h(x))$ is mean path length and $lrd_k$ is local reachability density. Features included hourly counts, 24h rolling z-scores, and time-of-day; the contamination rate (0.01) matches observed anomaly frequencies.


We specifically incorporated 24-hour rolling z-scores to capture local statistical deviations independent of absolute traffic volume. This feature allows the model to distinguish between a sudden, high-velocity `attack-like' burst and a gradual, legitimate increase in scientific workload. By coupling this with time-of-day encoding, the detector learns to suppress false positives during known peak operational hours, adapting the security baseline to the network's circadian rhythm.

We systematically test four ways to inject anomaly information: (1) raw (no anomaly), (2) mask-out, (3) score-as-feature, and (4) both combined. The general forecasting formulation shown in Eq.~\eqref{eq:forecast_formulation}:
\begin{equation}
\hat{y}_{t+1:t+h} = f(X_{t-w:t}, A_{t-w:t}, \theta)
\label{eq:forecast_formulation}
\end{equation}
where $X$ is historical traffic, $A$ is anomaly context, and $f$ is the forecasting model establishing the dynamic baseline.

Model benchmarking spanned 960 experiments (10 routers $\times$ 6 models $\times$ 4 horizons $\times$ 4 anomaly modes). All used chronological splits, early stopping, and cross-model normalization. Paired t-tests with Bonferroni correction, reported confidence intervals, and hardware control (NVIDIA A40 GPUs) ensure statistical rigor.

\section{Experimental Design for Security Baselines}

Building on our anomaly-aware pipeline, we now detail the experimental setup designed to stress-test these models as dynamic security baselines. To address the challenge of distinguishing legitimate scientific flows from volumetric attacks, we benchmark six representative model families, each selected to evaluate a specific property of baseline resilience: statistical rigidity, computational efficiency, or long-term temporal awareness.

\subsection{Model Portfolio and Security Rationale}

The design and selection of hybrid deep learning architectures is a critical factor in system performance across various domains \cite{shariff2025novel, hossain2020novel}. For this study, We selected six model families covering the spectrum of baseline generation techniques:

\begin{itemize}
    \item \textbf{SARIMA (Statistical Baseline):} Provides a rigid, seasonality-based baseline. Useful for testing whether simple statistical bounds are sufficient for REN security.
    \item \textbf{XGBoost (Tree Ensemble):} Excels at non-linear prediction with low inference latency, making it a candidate for high-speed edge security devices.
    \item \textbf{GRU-LSTM Hybrid (Recurrent DL):} Combines GRUs and LSTMs to capture stateful temporal dependencies, testing the value of ``memory" in distinguishing attacks from normal bursts.
    \item \textbf{N-BEATS (Interpretable DL):} Uses a stacked architecture for trend modeling. This model does not support exogenous features, allowing us to test its inherent architectural robustness without explicit anomaly scores.
   \item \textbf{TiDE (Time-series Dense Encoder):} An advanced MLP-based architecture that leverages dense encoder-decoder blocks with residual connections to learn long-range correlations efficiently without the overhead of self-attention mechanisms.
    \item \textbf{PatchTST (Patch-based Transformer):} Divides sequences into patches for computational efficiency. Our implementation does not support exogenous anomaly scores, serving as a control for `pure' Transformer performance.
\end{itemize}

This variability in feature support allows us to benchmark ``architecture-inherent robustness" versus ``explicit anomaly awareness," a critical distinction for security system design.

For our attention-based model (PatchTST), the core self-attention mechanism, which captures the long-range context critical for filtering out legitimate elephant flows, is formulated as shown in Eq.~\eqref{eq:attention}:
\begin{equation}
\mathrm{Attention}(Q,K,V) = \mathrm{softmax}\left(\frac{Q K^\mathsf{T}}{\sqrt{d_k}}\right)V
\label{eq:attention}
\end{equation}

\subsection{Experimental Protocol}

This factorial design explores how geographic diversity, model class, prediction horizon, and anomaly handling interact to affect the robustness of the security baseline. We employ a fully factorial design yielding 960 distinct experiments:

\begin{itemize}
    \item \textbf{Routers:} 10 geographically diverse Internet2 nodes.
    \item \textbf{Models:} The 6 families described above.
    \item \textbf{Horizons:} 1, 6, 12, and 24-hour prediction lead times, to evaluate baseline stability over operational shifts.
    \item \textbf{Anomaly Modes:} The four strategies (raw, mask-out, score-as-feature, and combined) to test resilience against data poisoning.
\end{itemize}

Data is split chronologically (70\% train, 15\% validation, 15\% test) to mirror a realistic security deployment where future attacks are unknown. All deep learning models were trained on NVIDIA A40 GPUs using early stopping (patience=20) to prevent overfitting to noise, with hyperparameters tuned via grid search (Table \ref{tab:hyperparams}).

\begin{table}[t!]
\centering
\caption{Key Hyperparameter Config determined via Grid Search}
\resizebox{\columnwidth}{!}{%
\begin{tabular}{@{}ll@{}}
\toprule
\textbf{Model} & \textbf{Key Settings} \\
\midrule
SARIMA     & $(1,1,1)\times(1,1,1)_{24}$ (AIC-optimized) \\
XGBoost    & n\_estimators=500, max\_depth=6, lr=0.05, early\_stop=50 \\
GRU–LSTM   & 128 units/layer, window=48h, 410 epochs, pat=20 \\
N-BEATS    & 30 stacks, 256 units/layer, 250 epochs \\
TiDE       & 1.6M params, window=48h, 130 epochs, lr=1e-3 \\
PatchTST   & 534K params, window=336h, patch\_len=16, stride=8 \\
\bottomrule
\end{tabular}
}
\label{tab:hyperparams}
\end{table}

\subsection{Evaluation Metrics for Baseline Fidelity}

We assess the fidelity of the security baseline using five complementary metrics, summarized in Table \ref{tab:evaluation_metrics}. Statistical significance is validated using paired t-tests ($\alpha=0.05$) with Bonferroni correction.

\begin{table}[h]
\centering
\caption{Evaluation Metrics for Baseline Fidelity}
\resizebox{\columnwidth}{!}{%
\begin{tabular}{@{}lll@{}}
\toprule
\textbf{Metric} & \textbf{Formula} & \textbf{Security Purpose} \\ \midrule
MAE & $\frac{1}{N}\sum |y_i - \hat{y}_i|$ & Avg. baseline deviation \\
RMSE & $\sqrt{\frac{1}{N}\sum (y_i - \hat{y}_i)^2}$ & Penalizes large deviations (potential attacks) \\
sMAPE & $\frac{100\%}{N}\sum \frac{|y_i - \hat{y}_i|}{(|y_i| + |\hat{y}_i|)/2}$ & Relative error for operator reporting \\
MSLE & $\frac{1}{N}\sum (\log(1+y_i) - \log(1+\hat{y}_i))^2$ & Robustness to legitimate volume spikes \\
P99 Error & 99th percentile of $|y_i - \hat{y}_i|$ & Critical for minimizing false positives during bursts \\ \bottomrule
\end{tabular}
}
\label{tab:evaluation_metrics}
\end{table}

\section{Results: Baselining and Resilience Analysis}

\subsection{Baseline Fidelity and Computational Cost}

\begin{figure}[h!]
    \centering
    \includegraphics[width=1\linewidth]{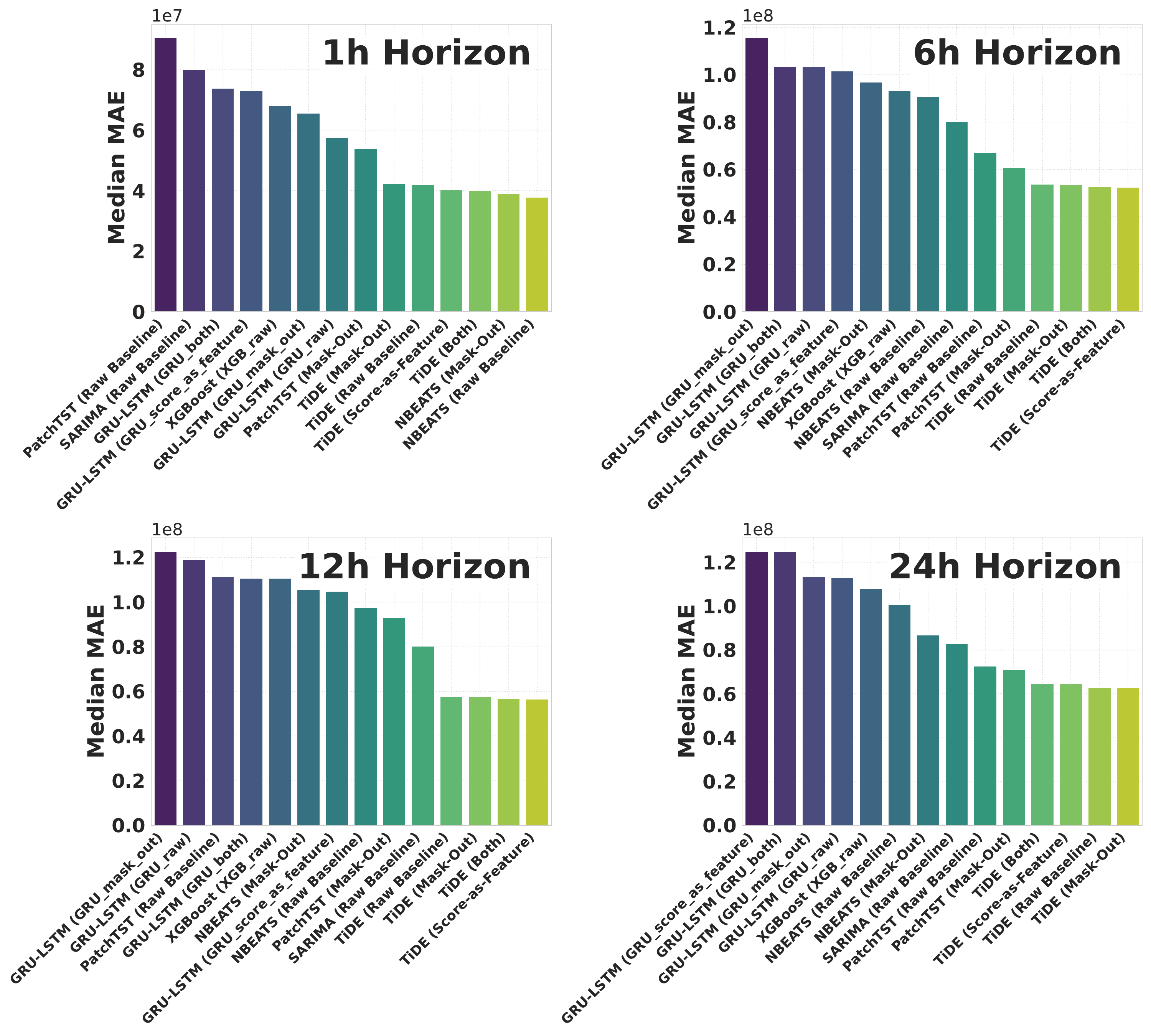}
    \caption{Baseline Fidelity (Median MAE) for all the Horizons. Transformer-based models (TiDE) provide the tightest security baseline, minimizing the residual noise that leads to false positives.}
    \label{fig:performance_summary}
\end{figure}

Our benchmarking reveals a clear hierarchy in the ability of models to establish a high-fidelity security baseline. As shown in Fig.~\ref{fig:performance_summary}, Advanced long-sequence architectures significantly outperform traditional methods, providing the tightest fit to normal traffic patterns and thus the lowest potential for false positives.

\textbf{Tier 1 – Dense Encoder Dominance}: TiDE variants achieve a median MAE of $5.66 \times 10^7$ packets, representing a 30–42\% error reduction over traditional baselines (Table~\ref{tab:performance}, $p<0.001$). For security operations, this superior fidelity is critical: it means legitimate bursts are correctly predicted as ``normal," preventing them from triggering false alarms in downstream IDS.

\textbf{Tier 2 - Modern Deep Learning}: N-BEATS and PatchTST form a strong middle tier. While they lack the exogenous anomaly integration of TiDE, their inherent architectural depth allows them to outperform RNNs by 25-30\%.

\textbf{Tier 3 - Lightweight Baselines}: XGBoost ($9.5 \times 10^7$ MAE) lags in accuracy but excels in computational efficiency. As shown in Table~\ref{tab:performance}, XGBoost training time (6.7s) and inference latency are negligible. This makes it a viable candidate for resource-constrained edge security appliances where ultra-low latency is prioritized over maximum baseline precision.

\begin{table}[!t]
\caption{Security Baseline Performance \& Cost (Median Values)}
\centering
\begin{tabular}{lrrrrr}
\toprule
\textbf{Model} & \textbf{MAE} & \textbf{RMSE} & \textbf{sMAPE} & \textbf{P99} & \textbf{Train (s)} \\
\midrule
TiDE (scF) & \textbf{5.66e7} & \textbf{7.5e7} & \textbf{60.4} & \textbf{1.9e8} & 1.6 \\
N-BEATS & 6.4e7 & 9.5e7 & 65.7 & 2.9e8 & 2.6 \\
PatchTST & 6.4e7 & 9.2e7 & 88.1 & 2.1e8 & 0.4 \\
GRU-LSTM & 8.8e7 & 1.3e8 & 87.1 & 3.9e8 & 1.8 \\
SARIMA & 8.1e7 & 1.2e8 & 122.5 & 2.9e8 & 1.9 \\
XGBoost & 9.5e7 & 1.1e8 & 86.4 & 3.4e8 & 6.7 \\
\bottomrule
\end{tabular}
\label{tab:performance}
\end{table}

\begin{figure}[h!]
    \centering
    \includegraphics[width=1\linewidth]{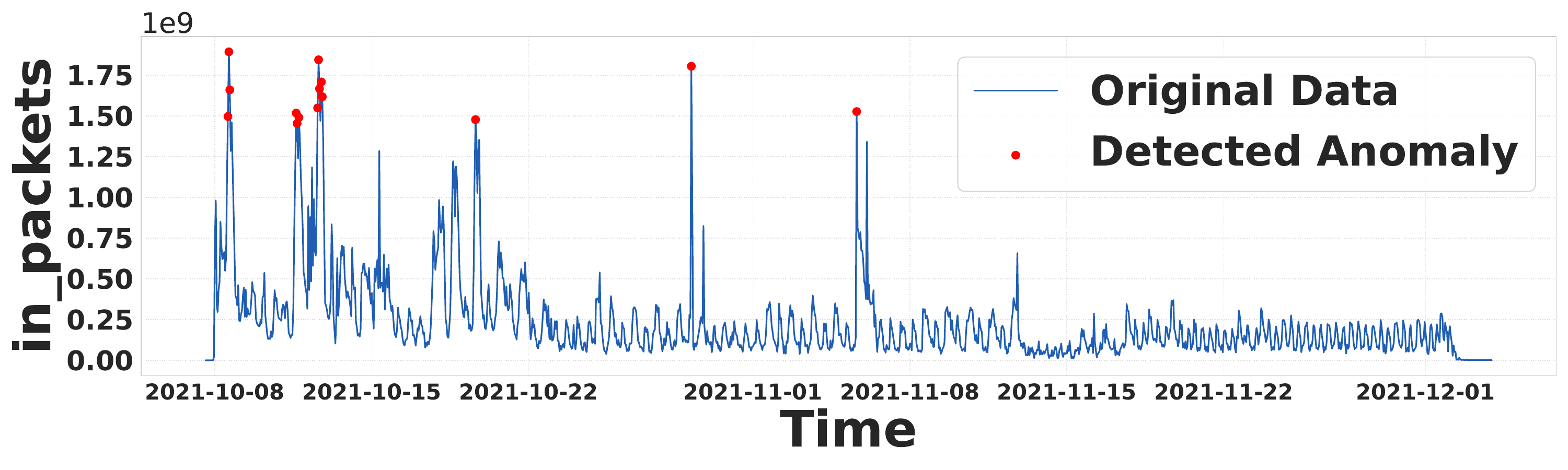}
    \caption{Visualizing the ``Noise": Legitimate traffic bursts (red dots) on the El Paso Router. A robust security model must predict these peaks to avoid flagging valid science flows as volumetric attacks.}
    \label{fig:anomaly_effects}
\end{figure}

Next, we analyze how robust these models are when trained on ``poisoned" data containing unflagged anomalies (Fig.~\ref{fig:anomaly_effects}).

\subsection{Resilience to Data Poisoning}

Security baselines must often be trained on historical data that contains unverified anomalies. Our ablation study (Table~\ref{tab:anomaly_impact}) measures each model's resilience to this ``training noise."

\textbf{RNN Vulnerability}: The GRU-LSTM model is highly sensitive to training noise. Masking anomalies (the ``Mask-out" strategy) yields a 3.3\% MAE reduction ($p < 0.01$), confirming that RNNs require clean training data to establish a stable security baseline.

\textbf{Dense Encoder Robustness}: TiDE benefits most from the ``Score-as-feature" strategy (1.11\% improvement, $p < 0.05$). This suggests that deep dense encoders can effectively use anomaly scores as a context signal to distinguish between normal high-volume transfers and statistical outliers.

\textbf{Inherent Stability}: Tree-based models (XGBoost) showed minimal sensitivity to anomaly integration strategies, indicating they are inherently robust to outliers which is a valuable trait for ``set-and-forget" security deployments.

\begin{table}[!t]
\caption{Resilience to Training Data Noise ($\Delta$MAE\% by Strategy)}
\centering
\begin{tabular}{lrrrr}
\toprule
\textbf{Model} & \textbf{Raw} & \textbf{Mask} & \textbf{Score} & \textbf{Both} \\
\midrule
GRU-LSTM & 0.0 & \textbf{-3.3***} & -0.7* & +3.4 \\
TiDE     & 0.0 & +0.06 & \textbf{-1.11**} & +0.86 \\
N-BEATS  & 0.0 & \textbf{-1.14**} & N/A\textsuperscript{a} & N/A\textsuperscript{a} \\
PatchTST & 0.0 & +5.89 & N/A\textsuperscript{b} & N/A\textsuperscript{b} \\
XGBoost  & 0.0 & -0.2 & -0.1 & +0.1 \\
\bottomrule
\end{tabular}
\raggedright\\
\footnotesize{Significance: * $p<0.05$, ** $p<0.01$, *** $p<0.001$} \\
\footnotesize{\textsuperscript{a}N-BEATS does not support exogenous features.} \\
\footnotesize{\textsuperscript{b}PatchTST implementation (used) did not support exogenous features.}
\label{tab:anomaly_impact}
\end{table}

\subsection{Operational Security Window}

The ``Operational Security Window" refers to how far in advance a security team can reliably predict traffic patterns to configure proactive defenses.

\begin{figure}[h!]
    \centering
    \includegraphics[width=0.8\linewidth]{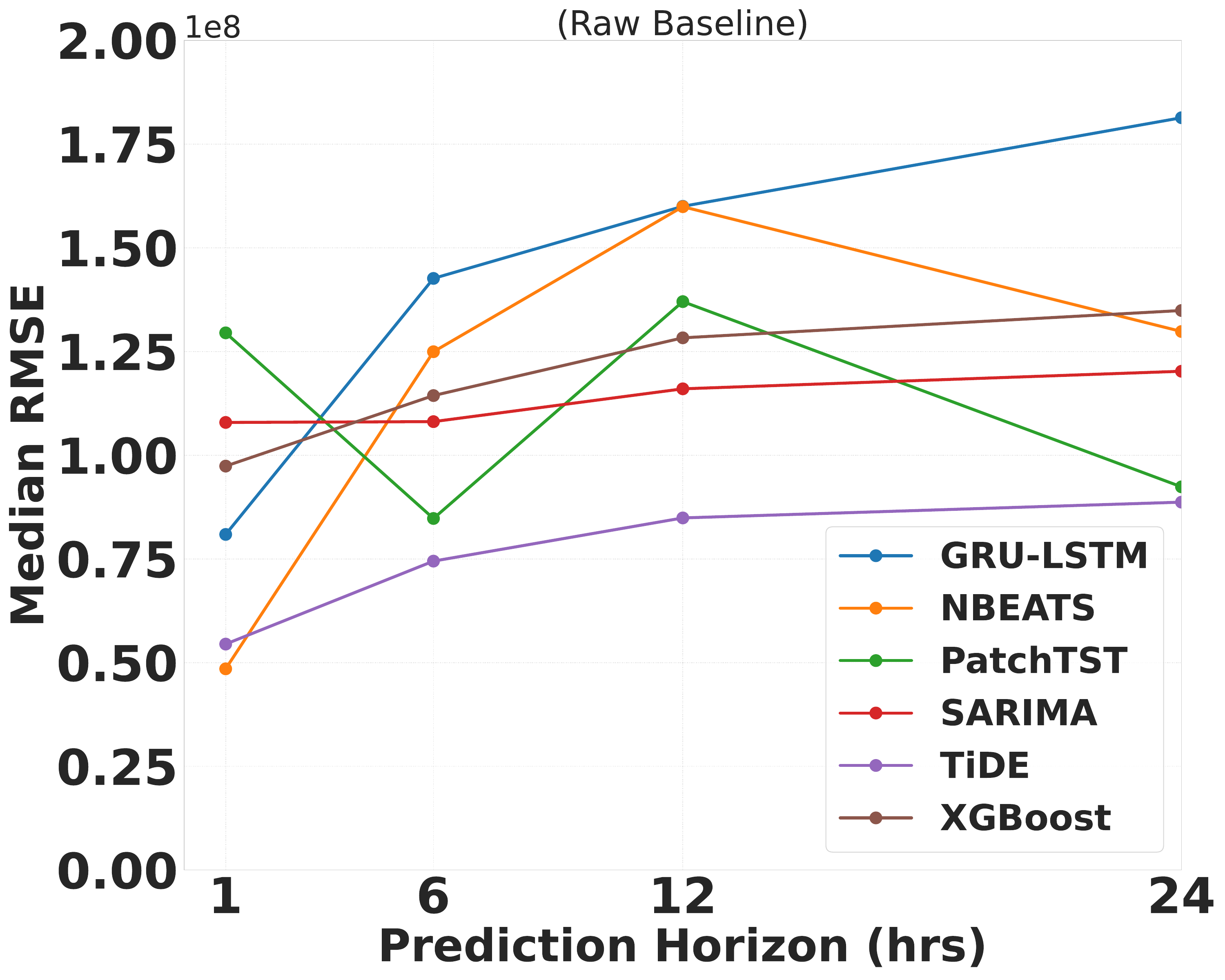}
    \caption{Operational Security Window. The sharp error increase up to 6 hours defines the optimal look-ahead window for proactive security planning and rule configuration.}
    \label{fig:lead_time}
\end{figure}

\textbf{1-6 Hour Window}: RMSE increases dramatically ($\approx$ 38\% median increase) in the short term. This reflects the difficulty of predicting the precise onset of scientific workflows. Security systems operating in this window must remain highly adaptive.

\textbf{6-24 Hour Stability}: Error growth flattens significantly after 6 hours. This identifies a 6-hour ``Sweet Spot" for security planning. Operators can reliably use 6-hour forecasts to differentiate expected elephant flows from sustained volumetric attacks, setting thresholds that are loose enough to accommodate science but tight enough to catch threats.

The baseline error growth follows a power law:
\begin{equation}
\text{MAE}(h) = 5.2 \times 10^7 \cdot h^{0.23} + 4.8 \times 10^7
\end{equation}
with $R^2 = 0.89$, providing a predictable error bound for security thresholds.

\subsection{Geographic Consistency of Baselines}

A global security policy requires baselines that work across diverse network segments. Our analysis reveals significant heterogeneity:

\begin{figure}[h!]
    \centering
    \includegraphics[width=1\linewidth]{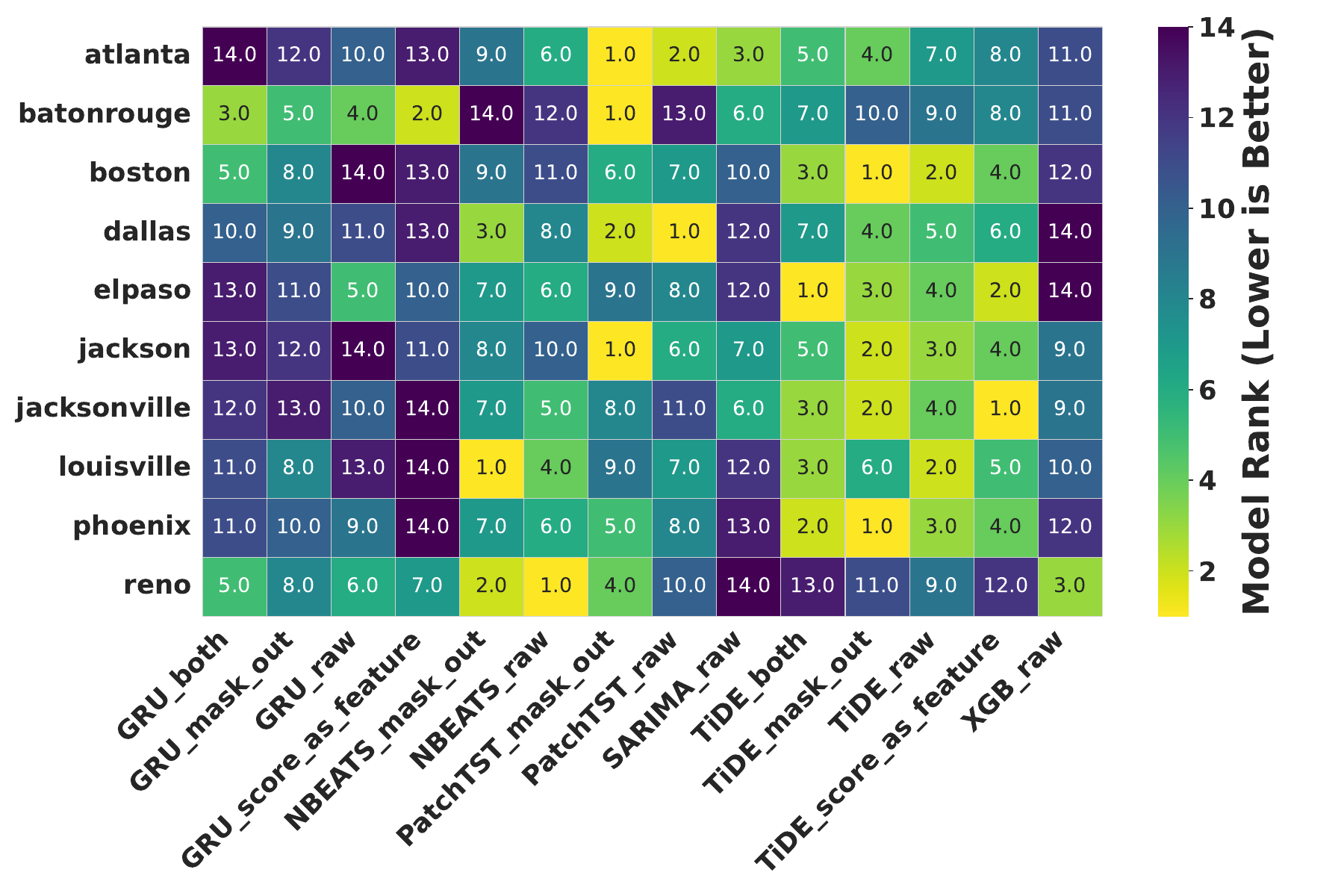}
    \caption{Baseline Consistency Heatmap by Router (24-hour MAE). Lower ranks (brighter colors) indicate better performance. The consistent high ranks for TiDE variants (top rows) across most routers demonstrate its robustness and suitability for network-wide security deployment.}
    \label{fig:router_performance}
\end{figure}

\textbf{High Predictability Nodes}: Regional nodes like Boston and Reno have stable traffic patterns, allowing for very tight security thresholds (low P99 error).

\textbf{High Variance Hubs}: Core hubs like Dallas and Jacksonville exhibit massive variance. Security baselines here must be looser to avoid false positives.

\textbf{Deployment Implication}: A ``one-size-fits-all" anomaly threshold will fail. Security teams must deploy per-router baselines, tuning the sensitivity based on the specific predictability profile of that node (as modeled by TiDE's consistent top-tier performance). While our current evaluation is limited to the Internet2 backbone, the ability of our framework to successfully model the extreme variance between regional nodes (e.g., Boston) and international exchange hubs (e.g., Dallas) provides strong evidence that the methodology will generalize to other REN environments with diverse topologies.

\section{Discussion, Conclusion, and Future Work}

Our results provide a clear verdict on baseline fidelity and deep insights into the architectural properties required for securing REN environments. Advanced long-sequence models demonstrate clear superiority. For PatchTST, this is attributable to self-attention (Eqn.~\ref{eq:attention}) processing sequences in parallel. For TiDE, superiority stems from its deep, multi-layer perceptron (MLP) encoder-decoder structure, which efficiently captures long-range dependencies and demonstrates inherent robustness to anomalies by mapping them into a dense latent space.

\subsection{Operational Security Recommendations}

Based on our comprehensive evaluation, we provide the first evidence-based deployment framework for REN security baselining, summarized in Table \ref{tab:deployment}.

\textbf{Security Strategy:} Our analysis confirms that a 6-hour prediction horizon offers the optimal balance between baseline accuracy and actionable lead time. This allows security teams to pre-configure firewalls and QoS policies for expected elephant flows. The need for per-router optimization is critical to avoid false positives in diverse network segments.

\textbf{Resilience \& Cost:} For resource-constrained edge devices, XGBoost serves as an excellent ``failover" option, providing competitive baselines with minimal latency. For core routers, TiDE offers the highest fidelity. A weekly retraining cadence is recommended to adapt the security baseline to the evolving scientific workload.

The broader impact is significant: by enabling autonomous, predictive baselining, this work allows security systems to filter out legitimate bursts, focusing analyst attention on true threats.

\begin{table}[t!]
\caption{Operational Security Deployment Recommendations}
\centering
\resizebox{\columnwidth}{!}{%
\begin{tabular}{@{}ll@{}}
\toprule
\textbf{Security Aspect} & \textbf{Recommendation} \\
\midrule
Primary Baseline Model & TiDE with anomaly score integration \\
Look-ahead Window & 6 hours (optimal fidelity-utility balance) \\
Training Data Hygiene & \textbf{Mask-out} anomalies for RNNs \\
& \textbf{Score-integration} for deep architectures (TiDE) \\
Edge vs. Core & XGBoost for Edge; TiDE for Core \\
Retraining Frequency & Weekly (to mitigate concept drift) \\
\bottomrule
\end{tabular}
}
\footnotesize{**This framework synthesizes our key findings into a set of actionable guidelines for deploying a robust security baseline in a production REN.}
\label{tab:deployment}
\end{table}

\subsection{Conclusion and Future Directions}

This work presents the first large-scale, anomaly-aware benchmark for traffic forecasting on the Internet2 backbone. By analyzing 13.7 billion packets across 10 routers, we have demonstrated that modern long-sequence architectures (like TiDE) can reduce baseline prediction error by over 30\%, significantly improving the ability of security systems to distinguish legitimate scientific bursts from volumetric attacks.

While our study establishes a strong foundation, we acknowledge several limitations that define clear avenues for future research. First, our 57-day temporal scope cannot capture long-term seasonal effects tied to annual academic cycles; extended longitudinal studies are needed. Second, our univariate, per-router approach serves as a necessary first line of defense, but ignores broader network-wide spatial dependencies. Finally, validation on other RENs (such as ESnet, GÉANT) is required to confirm generalizability. Due to data privacy restrictions, the raw NetFlow datasets cannot be publicly released. To mitigate this and ensure reproducibility, we commit to releasing all processing scripts, model training code, and the whole comprehensive evaluation report for all variations.

Future work will focus on addressing these limitations across three critical security frontiers:

\begin{itemize}

    \item \textbf{Adversarial Robustness \& Spatio-Temporal Modeling:} Evaluating how these baselines perform against sophisticated ``model poisoning" attacks, and extending the framework to incorporate Graph Neural Networks (GNNs) to model inter-router spatial dependencies.

    \item \textbf{Privacy-Preserving Federated Learning:} Developing protocols to allow multiple RENs to collaboratively train security models without sharing sensitive traffic data.

    \item \textbf{Explainable AI (XAI):} Integrating interpretability tools to help security analysts understand why a baseline deviation occurred, fostering trust in autonomous response systems.

\end{itemize}

Ultimately, this work enables a shift from reactive intrusion detection to proactive, predictive resilience, ensuring the security and availability of the digital infrastructure that powers modern science.

\section{Acknowledgments}
We would like to thank the Internet2 consortium for their support without which this work would not have been possible. This work is sponsored in part by National Science Foundation (NSF) grant (Award Number: OAC-2322369) and U.S. Department of Energy (DOE) grant (Award Number: DE-SC0024648).

\balance
\bibliographystyle{IEEEtran}
\bibliography{references.bib}

\end{document}